\documentclass[letter,twocolumn]{jpsj3}
\usepackage{graphicx}

\usepackage{txfonts}
\usepackage{color}

\newcommand{\InK}{InCu$_3$(OH)$_6$Cl$_3$}
\newcommand{\CdK}{CdCu$_3$(OH)$_6$(NO$_3$)$_2\cdot$H$_2$O}
\newcommand{\YCu}{YCu$_3$(OH)$_6$Cl$_3$}
\newcommand{\YYCu}{Y$_3$Cu$_9$(OH)$_{18}$[Cl$_8$(OH)]}
\newcommand{\YCOB}{YCu$_3$(OH)$_{6}$Br$_2$[Br$_{x}$(OH)$_{1-x}$]}
\newcommand{\Herb}{ZnCu$_3$(OH)$_6$Cl$_2$}

\title{Magnetic excitations in the 1/3 plateau state in \InK}
 
\author{Moyu Kato$^1$, Hiroyuki K. Yoshida$^1$, R. Kumar$^1$, Yoshihiko Ihara$^1$\thanks{yihara@phys.sci.hokudai.ac.jp}}
\inst{$^1$Department of Physics, Faculty of Science, Hokkaido University, Sapporo 060-0810, Japan} 

\abst{
  Magnetic dynamics in \InK\, was investigated from the NMR relaxation rate measurement. 
  In \InK, the magnetization isotherm shows a plateau at the 1/3 of full-saturation magnetization, characterizing the 1/3 plateau state. 
  As the 1/3 plateau state appears above 7 T upto 14 T, the microscopic magnetic properties were investigated with the NMR measurement in steady fields. 
  The temperature and field dependence of $1/T_1$ measurement reveals a gap in the magnetic excitation spectrum and its evolution with field in the 1/3 plateau state. 
  The field dependence of spin gap provides an important information to understand the microscopic origin of 1/3 plateau state in the kagome antiferromagnets.
}

\begin{document}
\maketitle

Quantum kagome antiferromagnets are key materials for exploring many-body states of quantum spins.
When magnetic moments with spin $S=1/2$ are arranged on the kagome network, the geometrically frustrated antiferromagnetic interactions between neighboring spins prevent magnetic long-range ordering (LRO). 
Fluctuating magnetic moments at low temperatures gain quantum nature, leading to an intriguing magnetic state such as quantum spin liquid. 
Theoretical models to describe the quantum spin liquid state have still been actively discussed between topological gapped $Z_2$ \cite{Fu-Science2015} and gapless $U(1)$ \cite{Han-Nature2012} models. 
In real systems, one of the best candidates for the quantum spin liquid is a perfect kagome antiferromagnet \Herb\,, which does not show the antiferromagnetic LRO down to 0.35 K.\cite{Shores-JACS2005, Helton-PRL2007, Mendels-PRL2007}
The low-temperature measurements are, however, difficult because the anti-site disorder between Zn$^{2+}$ and Cu$^{2+}$ ions creates orphan spins between kagome planes and generates parasitic magnetism on top of the intrinsic magnetism of kagome plane. \cite{Vries-PRL2008}
The Cu-based minerals \YCu\, and \YYCu, studied as the new candidates of spin-liquid realization \cite{Sun-JMC2016, Puphal-JMC2017,Hering-NPJ2022}, 
also have the similar issues due to the structural disorders \cite{ Sun-PRM2021, Chatterjee-PRB2023}.

\begin{figure}
\begin{center}
\includegraphics[width=8cm]{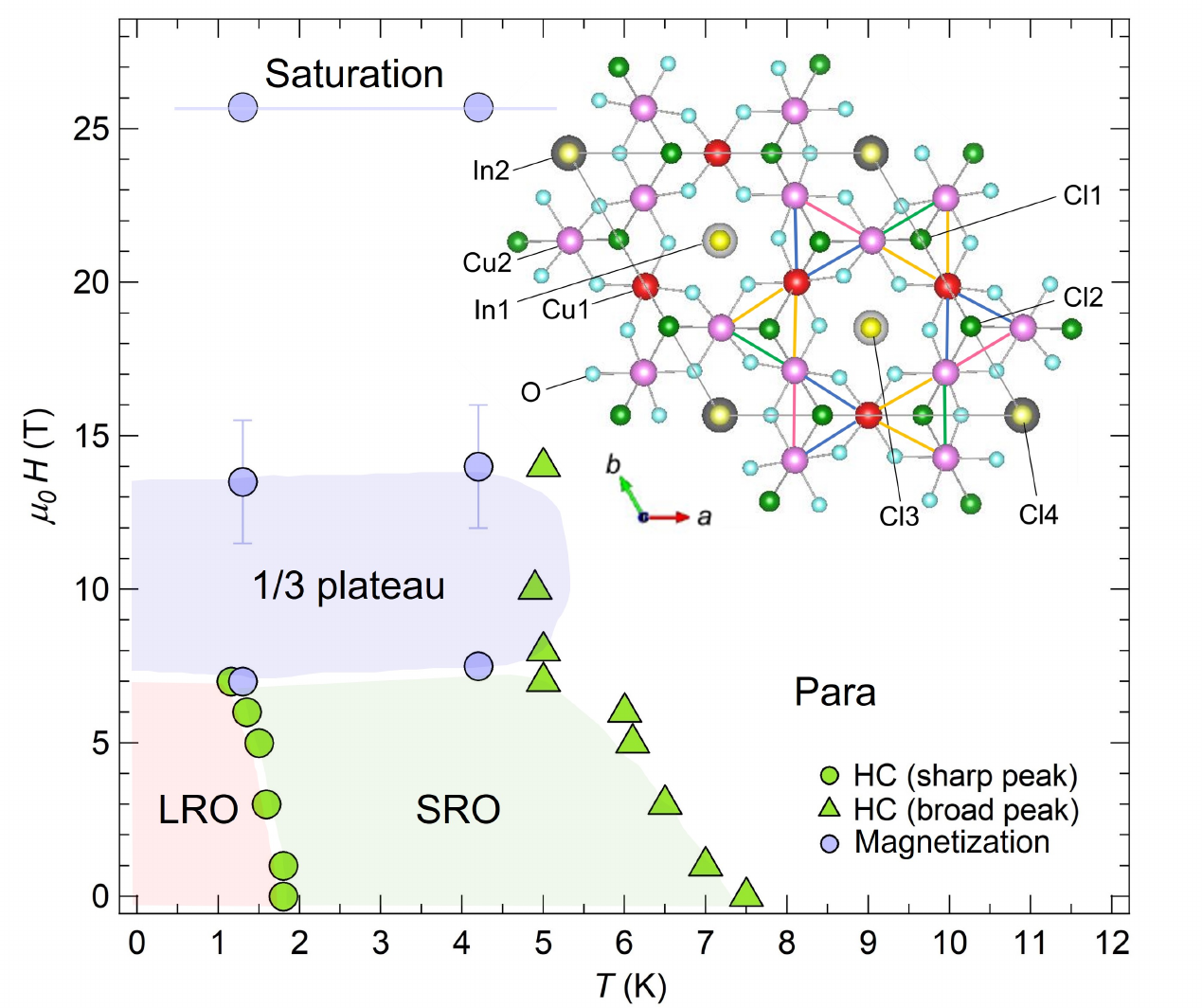}
\end{center}
\caption{
  Magnetic field-temperature phase diagram for \InK. \cite{Kato-CP2024} 
  The short-range order (SRO) and long-range order (LRO) phases were characterized by the heat capacity (HC) and magnetization measurements. 
  The magnetization isotherm measured with pulsed field shows a plateau behavior at 1/3 of full saturation between 7 T and 14 T before showing the full saturation above 25 T. 
  Inset shows the in-plane structure of \InK. 
}
\label{PhaseDiag}
\end{figure}

The intrinsic quantum state of kagome antiferromagnet has been further studied in high magnetic fields.
A small perturbation caused by various disorders becomes negligible with respect to the large energy scale of external fields.
The exotic magnetic states in fields are characterized by the magnetization plateau, which is a flat part of magnetization isotherm within a finite range of magnetic fields. 
The interesting magnetization plateau state has been observed recently in \YCOB\, around 20 T.  \cite{Jeon-NP2024, Suetsugu-PRL2024}
The magnetization plateau at 1/9 of full-saturation magnetization $M_{\rm sat}$ is theoretically explained as an exotic field-induced spin liquid state.
In addition to this, a Dirac spin liquid has been also suggested at zero field in \YCOB\, from the inelastic neutron scattering measurement. \cite{Chen-JMMM2020, Zeng-NP2024}
At higher magnetic fields, the successive magnetization plateaus were experimentally observed in \CdK \cite{Okuma-NC2019}.
The magnetization plateaus observed at quantized values of 1/3, 5/9, 7/9 of $M_{\rm sat}$ were explained by the crystallization of magnons excited around the hexagons of kagome network  \cite{Nishimoto-NC2013, Picot-PRB2016,Schulenburg-PRL2002} . 
The highly resonant quantum state robustly emerging in magnetic field is intriguing. 
In \YCOB, the onset of 1/3 plateau was observed above 50 T in addition to the 1/9 plateau, but the full saturation has not been reached as the exchange interaction $J$ is too strong. 
$J/k_B=65$ K for \YCOB\, gives an estimate of saturation field to be much higher than 100 T. 

Although active materials development has proposed several model materials with perfect kagome network, the magnetic properties up to the full saturation state have barely explored because extremely high magnetic fields are required to approach the field-induced magnetic states including 1/3 plateau state and microscopic measurements in such high field is challenging. 
The NMR spectroscopy is a powerful probe to investigate the microscopic properties of the field-induced states.
The nuclear spin-lattice relaxation rate $1/T_1$ measurement is essential to reveal the magnetic excitations by probing spin dynamics.   
For example, the temperature and/or field dependence of $1/T_1$ suggested a classical spin liquid state in $S=5/2$ kagome antiferromagnet Li$_9$Fe$_3$(P$_2$O$_7$)$_3$(PO$_4$)$_2$ \cite{Kermarrec-PRL2021} and a multi-magnon bound state in $S=1/2$ coupled trimer Cu$_3$V$_2$O$_7$(OH)$_2 \cdot$2H$_2$O.\cite{Yoshida-PRB2017}  
The relaxation time measurement can be performed even in high fields only accessible with pulsed field, when the time scale of $T_1$ is sufficiently short.\cite{Ihara-RSI2021,Kohama-JAP2022}
However, as $1/T_1$ measurement of long $T_1$ is still challenging in pulsed field, a kagome antiferromagnet with small $J$ is desired to investigate the novel quantum states.

To address the magnetic excitations in the 1/3 plateau state, we focus on the Cu-based mineral \InK. \cite{Kato-CP2024} 
The kapellasite-type structure of \InK\, has a nonmagnetic In$^{3+}$ at the center of kagome hexagon, as represented in the inset of Fig.~\ref{PhaseDiag}.
The crystal structure does not have the 6-fold symmetry but remaining 3-fold symmetry introduces the geometrical frustration and suppresses the long-range magnetic ordering temperature below 2 K.  
From the Curie-Weiss behavior of magnetization above 40 K, the nearest neighbor interaction was estimated to be $J_1/k_B=11.5$ K.  
The small $J_1$ provides us an easy access to the field-induced magnetic states.
In fact, magnetization measurement in pulsed field revealed a 1/3 plateau state at moderate magnetic fields from 7 T to 14 T and full saturation was achieved already at 26 T. \cite{Kato-CP2024} 
Figure \ref{PhaseDiag} summarizes the results of magnetization and heat capacity measurements in the magnetic field-temperature phase diagram.
In this study, we measured the temperature and field dependence of $1/T_1$ using the $^{35}$Cl-NMR signal and revealed the spin dynamics in and near the 1/3 plateau state.
Our results reveal a spin gap in the 1/3 plateau state and more importantly its field evolution around the border to the 1/3 plateau state.

The sample used for the NMR experiment was prepared through the hydrothermal synthesis. 
Cu chloride and In nitride as ingredients, lithium hydroxide as a catalyst were dissolved to the distilled water in a 25 ml Teflon-lined stainless steel autoclave, which was kept at 220 $^{\circ}$C for 3 days. 
Then, light blue polycrystalline samples were obtained as reported in Ref.~\cite{Kato-CP2024} .
\InK\, has a distorted kagome network formed by corner-shared isosceles triangles consisting of magnetic Cu$^{2+}$ ions.
The crystal structure (Space group $P31m$) has a 3-fold symmetry with the 3-fold axis on the nonmagnetic In ion sites located at the center of hexagon.
The Cl sites, our target of NMR measurement, are classified into two groups; 
Cl-1:$(0.3288,0.3288,0.8425)$ and Cl-2:$(0.6610,0.6610,0.1675)$ sites (site symmetry $m$) are located near the kagome plane connected to the Cu moments while Cl-3:$(1/3,2/3,0.0071)$ and Cl-4:$(0,0,0.9918)$ sites are not in the kagome planes and above the In sites at the center of hexagons. 
The site symmetry of Cl-3 and Cl-4 sites are $3$ and $3m$, respectively. 
For the NMR analyses, Cl-1 and Cl-2 sites on the mirror plane are labeled as Cl-m sites and Cl-3 and Cl-4 sites on the 3-fold axis are labeled as Cl-T sites.

For the $^{35}$Cl-NMR measurement, the randomly oriented powder sample was packed in a cylindrical sample cell. 
Its dimension is $\phi 3$ mm $ \times 5$ mm.
Knight shift measurements at high temperature were performed in a fixed magnetic field of approximately 13 T.
The field strength was calibrated by the $^{13}$C-NMR frequency of tetramethylsilane. 
At low temperatures, the NMR spectra were obtained by the field-sweep method, for which the NMR intensity at a fixed frequency was recorded during field sweep at a constant ramping rate.
The nuclear spin lattice relaxation rate $1/T_1$ was measured at the center line from the $m = 1/2 \leftrightarrow -1/2$ transition. 
The recovery profile of nuclear magnetization after the saturation pulse was recorded and fitted to the theoretical formula for the corresponding transition to estimate $T_1$.
A deviation from the theoretical curve was observed at low temperatures, but the fitting function for the single $T_1$ component was used for all the temperatures to consistently extract the average value of $T_1$.

\begin{figure}
\begin{center}
\includegraphics[width=7cm]{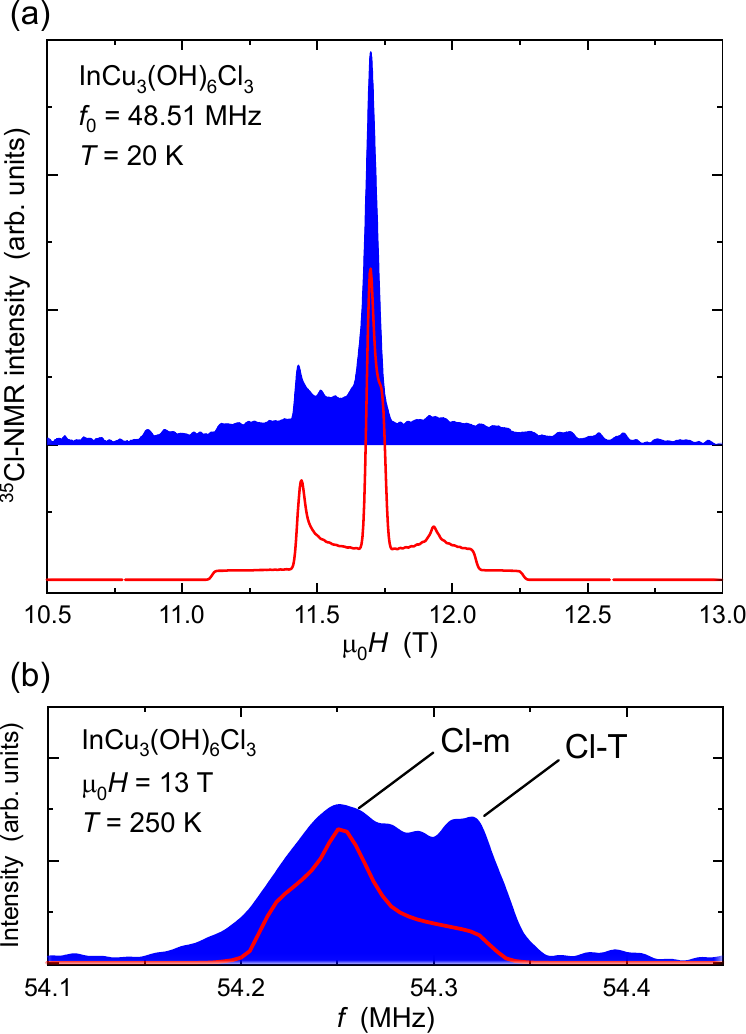}
\end{center}
\caption{
(a) Wide-range $^{35}$Cl-NMR spectrum measured at 20 K. 
The asymmetric spectral shape was consistently explained by the simulated powder spectrum represented by red solid line. 
(b) The high-resolution frequency spectrum for the center peak at 250 K. 
Red solid line is the simulated spectrum using the same parameters as in (a). 
The two peaks were assigned to the Cl-m and Cl-T sites. (See text)
}
\label{NMRsp}
\end{figure}

Figure \ref{NMRsp}(a) shows the full NMR spectrum for \InK\, obtained at 20 K and 48.51 MHz.
The $^{35}$Cl-NMR spectrum of powder sample is composed of a peak at the center and asymmetric structures at both sides. 
A small peak at 11.4 T and broad triangular structure around 11.9 T are consistently explained by introducing the asymmetry parameter of electric quadrupolar interaction and Knight shift. 
The simulated powder spectrum is shown by red solid line in Fig.~\ref{NMRsp}(a). 
From the parameters used for the simulation, the NQR frequency $\nu_Q$, the asymmetry parameter $\eta$ and asymmetric Knight shifts were estimated to be $\nu_Q = 2.40$ MHz, $\eta=0.15$, and $(K_{x}, K_y, K_z)  = (-1.1, -0.5, -0.5)$\%, respectively. 
The finite $\eta$ indicates that the observed $^{35}$Cl site does not have an axial symmetry. 
Here, $z$ axis of Knight shift is parallel to the main axis of the electric field gradient (EFG) at the Cl site.
The obtained $K_x<K_y=K_z$ relation suggests that the local $x$ axis is parallel to the crystallographic $c$ axis, meaning that the main axis of EFG is perpendicular to the $c$ axis. 
The axial symmetry around Cl-T sites should give $\eta=0$ and $K_{x} = K_y$, which is incompatible with the obtained asymmetric spectral shape.
Therefore, the observed $^{35}$Cl-NMR spectrum was assigned to the Cl-m sites, for which the local symmetry is lower than axial and the main EFG axis is within the kagome plane. 
The Cl-m sites cannot be further resolved into Cl-1 and Cl-2 sites as their local environment is similar. 
To detect the signal from Cl-T sites, the central peak was measured with better resolution at high temperature of 250 K, where the magnetic broadening is small, and the frequency spectrum is shown in Fig.~\ref{NMRsp}(b). 
The experimentally observed spectrum has two peaks.
The center peak of the powder spectrum was simulated using the parameter obtained from the full NMR spectrum to quantify the quadrupolar broadening and represented by red solid line in Fig.~\ref{NMRsp}(b). 
The spectral width is mainly explained by simulated powder pattern for Cl-m sites, except for the small peak at 54.31 MHz, which is assigned to the Cl-T sites, because the smaller site number of Cl-T should give smaller peak and the axial site symmetry results in narrower spectrum due to $\eta=0$. 
These two peaks are not clearly resolved by the magnetic broadening at low temperatures.
In the following, we measured the Knight shift $K$ and nuclear spin-lattice relaxation rate $1/T_1$ at the largest peak, which dominantly corresponds to Cl-m sites. 

\begin{figure}
\begin{center}
\includegraphics[width=8cm]{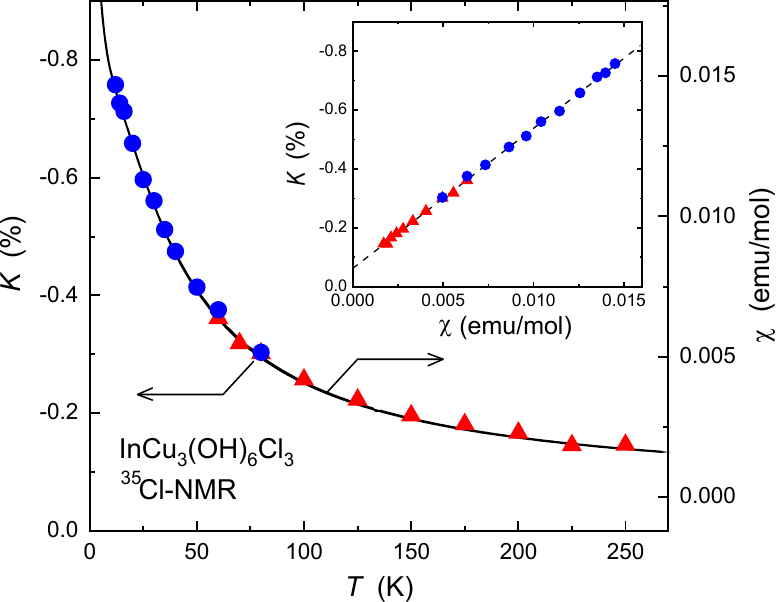}
\end{center}
\caption{
  Temperature dependence of Knight shift measured at 13 T (triangles) and 8.6 T (points). 
  The magnetization $\chi$ measured at 1 T is shown together with right axis. 
  Inset is the $K-\chi$ plot, in which $K$ is plotted as a function of $\chi$ using the temperature as an implicit parameter.  
}
\label{Kchi}
\end{figure}

The temperature dependence of Knight shift $K(T)$ is shown in Fig.~\ref{Kchi} by red triangles (13 T) and blue points (8.6 T).   
The spectra at low temperatures were obtained with the field-sweep measurement. 
The Curie-Weiss behavior of $K(T)$ is compared with the bulk magnetization $\chi(T)$ to estimate the hyperfine coupling constant $A$ through $K(T) = A \chi (T)$ relation.
The obtained $A =  -254$ mT/$\mu_B$ is negative due to the core-polarization mechanism for the hyperfine coupling. 
The negative and similar strength of coupling constants were reported for ZnCu$_3$(OH)$_6$Cl$_2$\cite{Kermarrec-PRB2014}, CaCu$_3$(OH)$_6$Cl$_2\cdot$0.6H$_2$O  \cite{Ihara-PRB2017}, and even in Br-NMR measurement for \YCOB\, \cite{Li-PRB2024}. 
At low temperatures, the characteristic two-peak structure evolves into more complex spectra below approximately 10 K, where a broad hump was observed in heat capacity.
However, the site assignment at low temperature is challenging for the powder spectrum with significant linewidth. 
The measurement for single crystal is required to resolve the low-temperature spectrum and investigate the magnetic structures therein. 

\begin{figure}
\begin{center}
\includegraphics[width=7cm]{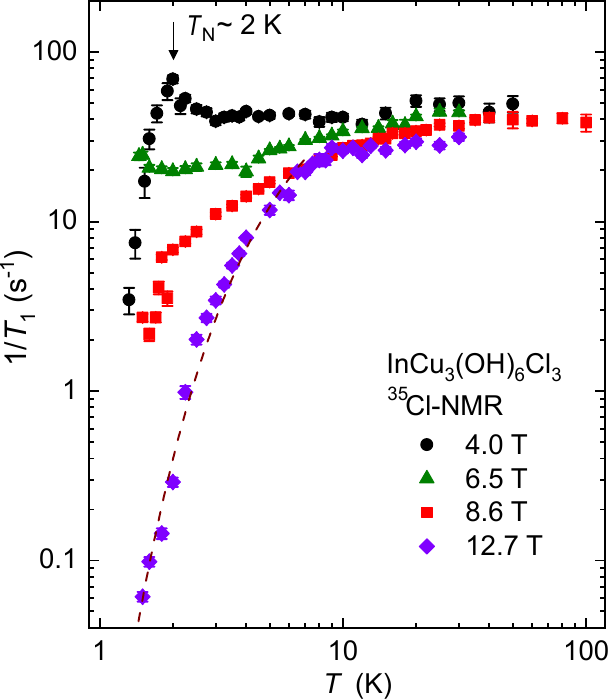}
\end{center}
\caption{
  Temperature dependence of $1/T_1$ measured at four field strengths crossing the boundary to the 1/3 plateau state. 
  At small field, a peak associated with LRO was observed at $T_{\rm N} \simeq 2$ K. 
  At higher fields, $1/T_1$ is suppressed below 10 K and finally exponential behavior was observed at 12.7 T in the 1/3 plateau state. 
  The dashed line is the result of fitting with the gap size $\Delta/k_B = 11$ K.
}
\label{T1}
\end{figure}

\begin{figure}
\begin{center}
\includegraphics[width=7cm]{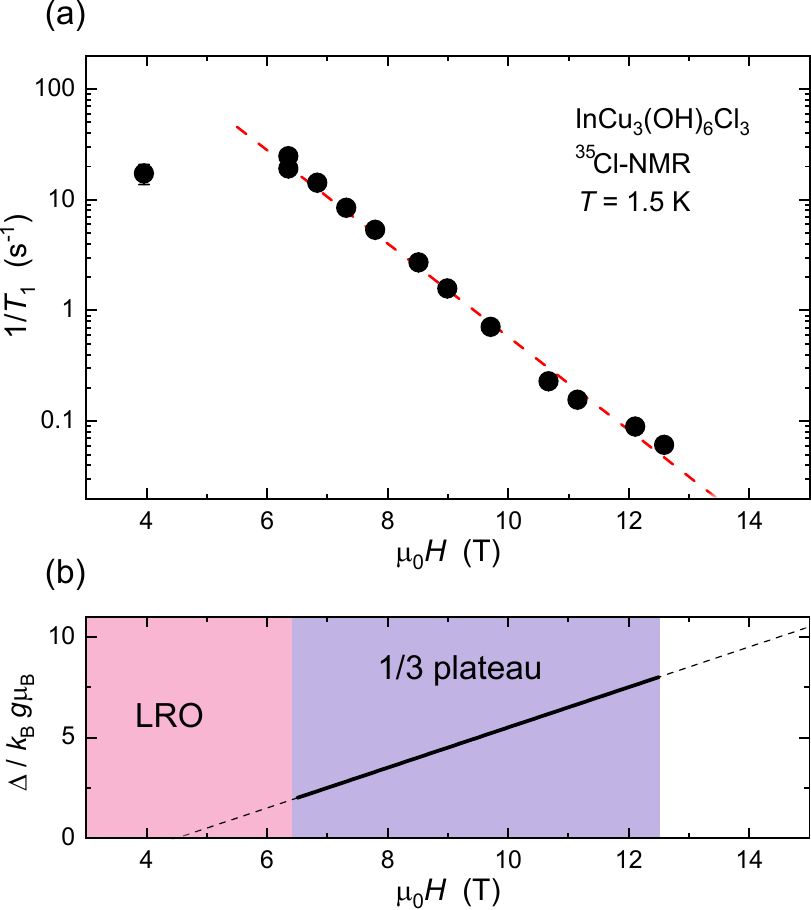}
\end{center}
\caption{
  (a) Field dependence of $1/T_1$ at 1.5 K. 
  An exponential behavior as a function of field was observed for broad field range covering the 1/3 plateau state. 
  A small value at low field was obtained in the LRO state. 
  (b) Field dependence of 1/3 plateau gap invoked from the $1/T_1$ result. 
  The gap appears around the onset field for the 1/3 plateau state and increase linearly with fields.
  }   
\label{gap}
\end{figure}

To explore the magnetic dynamics, the temperature dependence of $1/T_1$ was measured at four different fields across the onset of 1/3 plateau state.
At high temperatures above 50 K, $1/T_1$ is nearly independent of both temperature and field showing a constant value of $1/T_1 \simeq 40$ s$^{-1}$. 
This is a typical behavior for a localized spin system at temperature higher than the energy scale of exchange interaction. \cite{Moriya}
From the constant $1/T_1$ and the hyperfine coupling constant, the exchange interaction is estimated to be $J/k_B \simeq 5$ K, which is consistent with the bulk magnetization measurement. \cite{Kato-CP2024} 
In 4.0 T a peak was observed in the temperature dependence of $1/T_1$ at $T_{\rm N} \simeq 2$ K due to the critical magnetic fluctuations associated with LRO transition at low fields. 
The peak in $1/T_1$ is suppressed below 1.5 K by increasing fields up to 6.5 T and disappears in 8.6 T around the boundary of 1/3 plateau state.   
The field dependence of $T_{\rm N}$ is consistent with the previous study by heat capacity. \cite{Kato-CP2024}
When the LRO transition is suppressed by fields, the magnetic fluctuations above $T_{\rm N}$ is also suppressed as evident from the decrease in $1/T_1$ at 8.6 T starting below 10 K. 
The moderate temperature dependence in 8.6 T becomes steeper in the 1/3 plateau state at 12.7 T, 
where $1/T_1$ shows the thermal excitation-type exponential behavior. 
This result clearly suggests that the magnetic excitations are gapped in the 1/3 plateau state.
The gap size is estimated to be $\Delta/k_B = 11$ K by fitting the temperature dependence below 10 K with $1/T_1 = C\exp (-\Delta/k_BT)$.
The coefficient $C$ was also estimated to be $C=120$ s$^{-1}$. 
It is worth noting that the gap size is smaller than the Zeeman gap for an electron spin with $g \sim 2$ at 12.7 T. 
To reveal the origin of field-induced gap in the 1/3 plateau state, the detailed field dependence of $1/T_1$ was measured at 1.5 K crossing the boundary to the 1/3 plateau state.
Figure~\ref{gap}(a) shows that $1/T_1$ decreases exponentially with fields except for a small $1/T_1$ at 4.0 T. 
The deviation from the exponential behavior at low fields is ascribed to the LRO ground state. 
The exponential field dependence in the plateau state suggests that the gap in low-energy magnetic excitations evolves linearly with magnetic fields.  
By assuming a conventional Zeeman gap $\Delta(H) = g\mu_B\mu_0H$, $g$ factor is estimated from the slope in semi-log plot to be $g= 2.17$, which is consistent with the magnetization data \cite{Kato-CP2024} 
and excludes a possibility of multi-magnon process to dominate $1/T_1$ as found near the 1/3 plateau state in Cu$_3$V$_2$O$_7$(OH)$_2\cdot$2H$_2$O \cite{Yoshida-PRB2017} 
The absolute value of $1/T_1$ is, however, not explained by simply assuming that the Zeeman gap evolves from zero at zero field. 
To fit the experimental data as shown by red dashed line in Fig.~\ref{gap}(a), a field offset should be introduced, and thus, the field dependence of gap size becomes $\Delta(H) = g\mu_B(\mu_0H-4.5)$ as displayed in Fig.~\ref{gap}(b). 
Our result suggests that the finite magnetic gap appears around the phase boundary to the 1/3 plateau state and increases its size linearly with fields. 
The onset field of $\Delta(H)$ cannot be precisely determined from the present result limited above 1.5 K but is close to the onset of 1/3 plateau state. 

To summarize, we have measured the temperature and field dependence of $1/T_1$ around the 1/3 plateau state from the $^{35}$Cl-NMR measurement on the powder sample of \InK. 
The temperature dependence of $1/T_1$ in the 1/3 plateau state is characterized by the exponential behavior suggesting a spin gap in the magnetic excitation spectrum. 
To consistently explain the temperature and field dependence, we found that the spin gap appears around the onset of 1/3 plateau state and evolves linearly with field. 
The static magnetic structure in the 1/3 plateau state has not been investigated from the broad powder spectra.
To reveal the static magnetic configuration in the 1/3 plateau state, a single crystalline sample should be measured in future.

\begin{acknowledgements}
This work was partially supported by JSPS KAKENHI (Grants Nos. 21H01035, 22H04458, 22H00104, 23H04871, 25KJ0488). 
M.K. was supported by JST SPRING, Grant Number JPMJSP2119.
\end{acknowledgements}

\end{document}